\documentclass[tradictabstact]{aa}
\usepackage{graphicx}
\usepackage{natbib}
\bibpunct{(}{)}{;}{a}{}{,}
\usepackage{wasysym}

\usepackage{amsmath}
\usepackage{cancel}
\usepackage{multirow}
\usepackage{float}
\usepackage{color}

\begin{document}

\title{Penumbral thermal structure below the visible surface}
\author{J.M.~Borrero\inst{1} \and M.~Franz\inst{1} \and R.~Schlichenmaier\inst{1}
\and M.~Collados\inst{2,3} \and A.~Asensio Ramos\inst{2,3}}
\institute{Kiepenheuer-Institut f\"ur Sonnenphysik, Sch\"oneckstr. 6, D-79104, Freiburg, Germany
\and
Instituto de Astrof{\'\i}sica de Canarias, Avd. V{\'\i}a L\'actea s/n, E-38205, La Laguna, Tenerife, Spain
\and
Departamento de Astrof{\'\i}sica, Universidad de La Laguna, E-38205, La Laguna, Tenerife, Spain
}

\date{Recieved / Accepted}

\abstract{The thermal structure of the penumbra below its visible surface (i.e., $\tau_5 \ge 1$) has important implications
for our present understanding of sunspots and their penumbrae: their brightness and energy transport, mode conversion of magneto-acoustic waves,
sunspot seismology, and so forth.}{We aim at determining the thermal stratification in the layers immediately beneath the visible 
surface of the penumbra: $\tau_5 \in [1,3]$ ($\approx 70-80$ km below the visible continuum-forming layer)}{We analyzed spectropolarimetric 
data (i.e., Stokes profiles) in three Fe \textsc{i} lines located at 1565 nm observed with the GRIS instrument attached to the 1.5-meter solar 
telescope GREGOR. The data are corrected for the smearing effects of wide-angle scattered light and then subjected to an inversion code for 
the radiative transfer equation in order to retrieve, among others, the temperature as a function of optical depth $T(\tau_5)$.}
{We find that the temperature gradient below the visible surface of the penumbra is smaller than in the quiet Sun. This implies that in the region $\tau_5 \ge 1$ 
the penumbral temperature diverges from that of the quiet Sun. The same result is obtained when focusing only on the thermal structure below the surface 
of bright penumbral filaments. We interpret these results as evidence of a thick penumbra, whereby the magnetopause is not located near its visible 
surface. In addition, we find that the temperature gradient in bright penumbral filaments is lower than in granules. This can be explained in terms of the limited 
expansion of a hot upflow inside a penumbral filament relative to a granular upflow, as magnetic pressure and tension forces from the surrounding penumbral
magnetic field hinder an expansion like this.}{}

\titlerunning{Penumbral thermal structure below the visible surface}
\authorrunning{Borrero et al.}
\maketitle

\def\kms{~km s$^{-1}$}
\def\deg{^{\circ}}
\def\df{{\rm d}}
\newcommand{\ve}[1]{{\rm\bf {#1}}}
\newcommand{\diff}{{\rm d}}
\newcommand{\Conv}{\mathop{\scalebox{1.5}{\raisebox{-0.2ex}{$\ast$}}}}%

\section{Introduction}
\label{section:intro}

The thermal structure of sunspots, in particular below the visible surface, is of great importance for a number of scientific studies. It determines the sound
speed and therefore the inference of the subsurface structure of sunspots through the calculation of travel times through local helioseismology 
\citep[see ][ and references therein]{couvidat2006seismo,moradi2010seismo,gizon2009seismo,khomenko2015waves}. It also establishes the location of the
$\beta=1$-surface (i.e., the region where magnetic and gas pressure are equal) where most of the mode conversion of magnetoacoustic waves takes place 
\citep{cally2005mode,schunker2006mode,khomenko2012waves}. Furthermore, a proper knowledge of the thermal stratification can also be employed to assess 
the realism of three-dimensional magnetohydrodynamic simulations of sunspots \citep{heinemann2007mhd,rempel2009mhd,rempel2012mhd}. It also imposes 
strong observational constraints on the different models that aim at explaining the onset of convection and heat transport in sunspot penumbrae: 
convective rolls \citep{danielson1961pen}, flows along magnetic  flux tubes \citep{schliche1999pen}, or overturning convection in field-free gaps 
\citep{spruit2006gap,scharmer2006gap}. Finally, it can help elucidate whether the subsurface structure of the penumbra is better explained in terms 
of the monolithic \citep{deinzer1965sunspot,meyer1974sunspot} or cluster/spaghetti \citep{parker1979sunspot} models.\\

Unfortunately, the subsurface layers are not easily accessible. For instance, the analysis of spectral lines and their polarization 
signals (i.e., spectropolarimetry) allows inferring the physical properties of solar and stellar atmospheres
\citep{chandra1960book,deltoro2003book} as a function of the continuum optical depth $\tau_c$, where $\tau_c=1$ is considered
the deepest observable layer and is oftentimes referred to as the continuum-forming layer. However, the wavelength dependence of the 
continuum optical depth offers the possibility of observing slightly deeper or higher layers of the solar atmosphere by observing different 
wavelengths \citep{chandra1946hminus}. The observed surface of the Sun  \citep[see Figure~9
in][]{borrero2016gregor} is located 
about $70-80$ km deeper at 1565 nm (near-infrared) than at 630 nm (visible): $z(\tau_{15}=1)-z(\tau_{6}=1) \approx 70-80$ km. Here, the
subindexes $15$ and $6$ indicate wavelengths of 1565 nm and 630 nm, respectively. If instead of using two different reference wavelengths
we employ only the commonly used reference wavelength of 500 nm, it can be shown \citep[see Figure~10 in][]{borrero2016gregor} that the response
function to the temperature for spectral lines at 630 nm peaks at $\tau_5 \approx 1$, while for spectral lines at 1565 nm, it peaks at around
$\tau_5 \approx 3$. In this work we exploit this property to investigate the penumbral thermal stratification $T(\tau_5)$ in the optical depth 
range $\tau_5 \in [1,3]$, that is, in the first $\approx 70-80$ km below its visible surface.\\

\section{Observations and inversion results}
\label{section:obs}

The observations employed in this work correspond to spectropolarimetric data (i.e., the Stokes vector as a function of 
wavelength $\ve{I}(\lambda)=(I,Q,U,V)$) in three Fe \textsc{i} lines at 1565 nm. Although the spectral region contains 
six neutral iron spectral lines with a spectral resolution of $40$ m{\AA} pix$^{-1}$, here we analyze only the three 
with the largest Land\'e factor and cleanest continuum: Fe \textsc{i} 1564.8 nm, Fe \textsc{i} 1565.2 nm, and Fe 
\textsc{i} 1566.2 nm. The atomic parameters of these spectral lines can be found in Table 1 in \cite{borrero2016gregor}.
The data were acquired with the GREGOR Infrared Spectrograph \citep[GRIS; ][]{collados2012gregor} coupled 
with the Tenerife Infrared Polarimeter \citep[TIP2; ][]{collados2007tip} that is attached to the 1.5-meter 
GREGOR telescope \citep{schmidt2012gregor}. The target, NOAA 12049, was observed on May 3, 2014, very 
close to disk center (heliocentric angle $\Theta=6.5^{\deg}$). All details of the data reduction and post-processing can be found in
\cite{borrero2016gregor,morten2016gregor,lagg2016gregor,marian2016gregor}. In addition, a PCA-based
deconvolution \citep[principal component analysis; see][]{basilio2013pen} was performed in order to account for an estimated 20 \% of 
wide-angle scattered light in the observations. The noise in the final data is about $10^{-3}$ (in units of the 
quiet-Sun continuum intensity), and the spatial resolution is estimated to be about 0.4-0.45\arcsec.\\

\begin{figure*}
\begin{center}
\includegraphics[width=14cm]{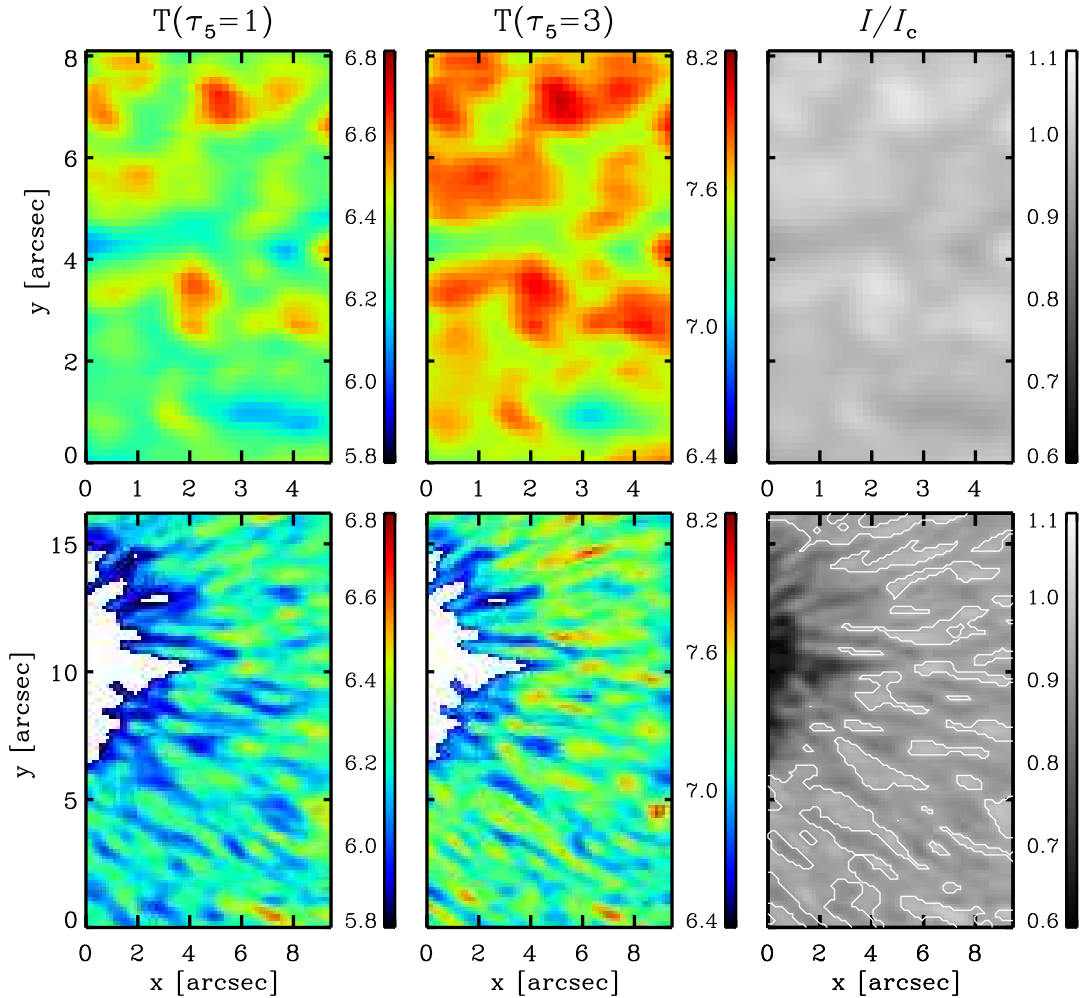}
\caption{Temperatures in two different regions: quiet Sun (top) and penumbra (bottom). Results at optical depths $\tau_5=1$ and $\tau_5=3$ are displayed 
in the left and middle panels, respectively. Right panels show the normalized continuum intensity images. White contours in the bottom right panel 
(continuum intensity in the penumbra) enclose bright penumbral filaments, whereas white regions in the penumbral temperature maps (left and middle bottom panels) 
indicate the umbra. All values of the temperature are given in thousands of Kelvin.\label{figure:temperaturesic}}
\end{center}
\end{figure*}

While the size of the scanned region is 53$\times$45 arcsec$^2$ \citep[see bottom panel in Fig.~1 in][]{borrero2016gregor},
in this work we focus on two distinct smaller regions. The first region is a quiet-Sun patch with an area of 4.9$\times$8.3 arcsec$^{2}$
that contains about 2200 pixels. The second region is a 9.6$\times$16.5 arcsec$^{2}$ penumbral patch located on
the limb side that contains about 8600 pixels. The right panels in Figure~\ref{figure:temperaturesic} display the normalized
continuum intensity in these regions. The normalization is done such that the average continuum intensity
in the quiet-Sun region is unity.\\

The PCA-deconvolved spectropolarimetric data have been analyzed with the SIR \citep{basilio1992sir} inversion 
code in order to retrieve the physical parameters (temperature, magnetic field, line-of-sight velocity, etc.) 
as a function of optical depth $\tau_5$. Owing to the removal of scattered light, we consider only 
one atmospheric component within each pixel \citep{borrero2016gregor}. In order to minimize the probability of arriving at a local minimum, the 
inversion at each pixel has been repeated ten times employing randomly generated initial guess models. These initial models feature
$\tau$-independent values (randomly generated using a uniform distribution of values) in the kinematic and magnetic parameters. The thermal
stratification is generated by performing random perturbations (also using a uniform distribution) at different optical depth locations
on the HSRA model \citep{hsra1974} and reinterpolating $T(\tau_5)$ in between these locations. Out of the ten inversion results we kept only 
the solution with the smaller $\chi^2$. Because in this work we are mostly interested in the temperature stratification $T(\tau_5)$, which is mostly 
encoded in Stokes $I$, we gave all four Stokes parameters equal weight in the inversion $w_{i,q,u,v}=1$ \citep[cf.][~]{borrero2016gregor}. 
During the inversion, the temperature was allowed to change in three nodes, while the components of the magnetic field $(B,\gamma,\phi)$ and the 
line-of-sight velocity $v_{\rm los}$ were allowed to change in two nodes when inverting the penumbral area, but only one node in the quiet-Sun area.\\

Figure~\ref{figure:temperaturesic} shows the resulting temperatures at two optical depths $\tau_5=1$ (left) and
$\tau_5=3$ (middle) in the quiet-Sun (upper panels) and penumbral regions (bottom panels). The temperatures clearly closely follow the continuum intensity maps in the right panels of the same figure. For visualization
purposes, the umbral region, defined as those regions where the continuum intensity is lower than 0.85, is drawn in white
on the temperature maps. Regions where the temperature is below the lower limit of the color bars are also shown 
in white. On the other hand, white contours in the continuum intensity map of the penumbra (bottom right panel) indicate
the location of \textup{bright penumbral }filaments. Pixels belonging to bright filaments are selected as those penumbral pixels 
where the continuum intensity is above the mean continuum intensity at the same radial distance from the sunspot center
as the pixel subjected to scrutiny. As expected, the temperature in these bright filaments is generally higher than in the dark 
penumbral regions. Table~\ref{table:temptau} summarizes the mean temperatures at optical depths $\tau_5=1$ and $3$, as well as their
difference, in different regions: average quiet Sun, granules only, intergranules only, average penumbra, and bright penumbral filaments.\\

\begin{table}
\begin{center}
\caption{Mean temperatures (in Kelvin) at two different optical depths and their difference.\label{table:temptau}}
\begin{tabular}{cccc}
Region & $\overline{T}(\tau_5=1) $ & $\overline{T}(\tau_5=3)$ & $\Delta \overline{T}$ \\
\hline
Quiet Sun & $6338 [\pm 67] $ & $7628 [\pm 80]$ & $1290 [\pm 104]$ \\
Granules & $6412 [\pm 67]$ & $7756 [\pm 80]$ & $1344 [\pm 104]$  \\
Intergranules & $6266 [\pm 67]$ & $7502 [\pm 80]$ & $1236 [\pm 104]$ \\
Penumbra & $5964 [\pm 67]$ & $6927 [\pm 80]$ & $963 [\pm 104]$ \\
Bright filaments & $6081 [\pm 67]$ & $7138 [\pm 80]$ & $1057 [\pm 104]$ \\
\end{tabular}
\end{center}
\tablefoot{Errors have been estimated through a Monte Carlo-like simulation of Stokes profiles synthesized
from three-dimensional magnetohydrodynamic simulations \citep{rempel2011mhd,rempel2012mhd} that were inverted in 
the same fashion as described in Sect.~\ref{section:obs} after adding photon noise to a level of $10^{-3}$.}
\end{table}

\section{Discussion and interpretation}
\label{section:dis}

\subsection{Penumbra and bright filaments vs average quiet Sun}
\label{subsection:xidis}

\begin{figure}
\begin{center}
\includegraphics[width=8cm]{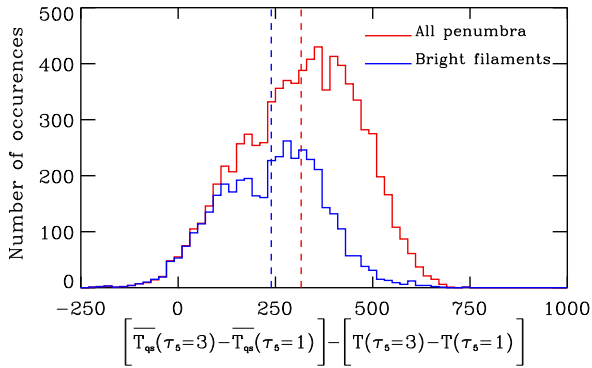}
\caption{Histograms of $\xi(T)$ as defined in Equation~\ref{equation:deltat}. Red lines correspond to the penumbral region (bottom 
panels in Fig.~\ref{figure:temperaturesic}). Blue lines correspond to the bright penumbral filaments 
(white contours in the bottom panels of Fig.~\ref{figure:temperaturesic}). The vertical dashed lines indicate the mean
values of the histograms, $\overline{\xi(T)}$.\label{figure:deltathist}}
\end{center}
\end{figure}

A particularly interesting feature is the finding that the difference of the mean temperatures between the quiet Sun and the 
penumbra is larger at $\tau_5=3$ than at $\tau_5=1$ (see Table~\ref{table:temptau}). At $\tau_5=1,$ the average quiet Sun is about $(374 \pm 95)$ 
K hotter than the average penumbra, while this difference increases to $(701 \pm 113$ K at $\tau_5=3$. This also occurs when comparing only the bright penumbral filaments 
to the quiet Sun:  at $\tau_5=1,$ the average quiet Sun is about $(257 \pm 95)$ K hotter than the average over the bright filaments, while this difference 
increases to $(490 \pm 113)$ K at $\tau_5=3$. This feature is not only seen in the average temperatures, but  also in the individual temperatures inferred at 
each pixel. Figure~\ref{figure:deltathist} shows the histograms of the difference between the mean quiet-Sun temperature at $\tau_5=3$ and $\tau_5=1$, 
minus the same difference at each pixel of the penumbra (red) or at each pixel of the bright filaments (blue). Hereafter, this quantity is
referred to as $\xi(T)$,

\begin{eqnarray}
\begin{split}
\xi(T) = & \big\{\overline{T_{\rm qs}}(\tau_5=3)-\overline{T_{\rm qs}}(\tau_5=1)\big\}- \\ &
\big\{T(\tau_5=3)\big\}-T(\tau_5=1)\big\} \;.
\label{equation:deltat}
\end{split}
\end{eqnarray}

Locations where $\xi(T)<0$ are regions where the temperature stratification $T(\tau_5)$ converges 
to the mean quiet-Sun stratification as we probe below the visible surface ($\tau_5\in [1,3]$ or about $70-80$ 
km below the surface), while pixels where $\xi(T)>0$ correspond to regions where $T(\tau_5)$ diverges from 
the mean quiet-Sun stratification. The determination of the temperature in this region has been possible because 
the analyzed spectral lines (i.e., three Fe \textsc{i} lines located at 1565 nm) convey reliable information about 
the temperature at this depth. A similar study employing data from Hinode/SP is not possible because of the 
large uncertainties in the retrieval of the temperature at $\tau_5=3$ using the Fe \textsc{i} spectral lines at 630 nm.\\

Most of the penumbra and bright filaments are formed by pixels where $\xi(T)>0$, with pixels where $\xi(T)<0$
representing only 1.3 \% and 2.8 \% of the total, respectively. As in \cite{borrero2016gregor}, we have checked that 
the results we present here are independent of the amount of wide-angle scattered light considered during the deconvolution 
process: the very same results are obtained when considering $40$ \% of the wide-angle scattered light as well as when analyzing 
the original (i.e., not deconvolved) Stokes vector.\\

The vertical dashed lines in Fig.~\ref{figure:deltathist} indicate the mean value of the histograms. We refer to it
as $\overline{\xi(T)}$. They are located at at $\overline{\xi(T)}=(327 \pm 147)$~K for the entire penumbra (dashed red) and at 
$\overline{\xi(T)}=(233 \pm 147)$~K for the bright penumbral filaments (dashed-blue).

We have found that both on average and in 98.7 \% of the studied penumbral area, $\xi(T)>0$, 
which implies that the temperature gradient below the visible surface of the penumbra is lower than the 
mean temperature gradient of the quiet Sun. A particular interpretation that can be drawn 
from these results is related to the penumbral thickness below the visible surface. 
The concepts of thin and thick penumbrae have been discussed on theoretical grounds 
\citep[see, e.g., ][]{jahn1994sun,solanki2003review}, but so far it was not possible to distinguish 
them observationally. While in a thin penumbra the magnetopause (defined as the current sheet separating the 
penumbral magnetic field from the surrounding convection) is located at $\tau_5 \approx 1$, in a thick penumbra it 
lies at greater depth. Therefore, if the penumbra were thin, the temperature between $\tau_5 \in [1,3]$ would converge toward quiet -Sun values, 
and therefore feature $\xi(T)<0$. However, our results show that the opposite occurs: $\xi(T)>0$, thereby favoring a thick penumbra.\\

It has also been suggested that while in the penumbral spines (regions of stronger and less strongly inclined magnetic field) 
the magnetic field reaches farther down, in the penumbral intraspines (regions of weaker and more inclined magnetic field) the magnetic 
field quickly yields to field-free convection below $\tau_5=1$ \citep[see Fig.~3 in][]{spruit2006gap}. Interestingly, 
this latter possibility can be ruled out because the high spatial resolution ($\approx 0.4-0.45"$) spectropolarimetric data
of GREGOR\ allow us to establish that $\xi(T)>0$ also below bright penumbral filaments (i.e., intraspines), both on average
and in 97.2 \% of the area covered by bright filaments.\\

\subsection{Bright filaments vs granules}
\label{subsection:upflowdis}

Notwithstanding the previous results about the penumbral thickness, there is compelling evidence for the existence of 
convective-like motions inside the magnetized bright penumbral filaments \citep{zakharov2008pen,rempel2012mhd,tiwari2013decon}, 
in particular of some form of convective upflow at their center. In the optically thick region we are studying ($\tau_5 \ge 1$), the thermal
stratification $T(\tau_5)$ is dominated by the adiabatic expansion (i.e., cooling) of the plasma as it rises. Because of this, we
might surmise that the 
temperature gradient in the quiet-Sun granules and bright penumbral filaments would be very similar. Our results indicate, however, that the 
opposite occurs (see Table~\ref{table:temptau}). A possible solution to this seemingly contradictory results can be offered by considering that
while in the quiet Sun this adiabatic expansion can take place in all spatial directions, in bright penumbral filaments (i.e., intraspines) it will be 
limited by the magnetic pressure and tension of the external field (i.e., spines). Consequently, the expansion of the magnetized upflow inside the bright 
filament will have fewer degrees of freedom, thereby leading to a decreased cooling and thus the smaller temperature gradient observed.\\

\begin{acknowledgements}
The 1.5-meter GREGOR solar telescope was built by a German consortium under the
leadership of the Kiepenheuer-Institut f\"ur Sonnenphysik in Freiburg with the
Leibniz-Institut f\"ur Astrophysik Potsdam, the Institut f\"ur Astrophysik
G\"ottingen, and the Max-Planck-Institut f\"ur Sonnensystemforschung in G\"ottingen as
partners, and with contributions by the Instituto de Astrofísica de Canarias and
the Astronomical Institute of the Academy of Sciences of the Czech Republic. Financial support 
by the Spanish Ministry of Economy and Competitiveness through projects AYA2014-60476-P 
and Consolider-Ingenio 2010 CSD2009-00038 are gratefully acknowledged. We would like to
thank Matthias Rempel, Fernando Moreno-Insertis, and Oskar Steiner for fruitful discussions.
This research has made use of NASA's Astrophysics Data System.
\end{acknowledgements}

\bibliographystyle{aa}
\bibliography{ms}

\end{document}